\documentclass[runningheads]{llncs}

\usepackage{layouts}

\usepackage[T1]{fontenc}
\usepackage{graphicx}
\usepackage{amssymb}
\usepackage{amsmath}
\usepackage{bm, bbm}
\usepackage{caption}
\usepackage[capitalize]{cleveref}
\usepackage[inline]{enumitem}
\usepackage{mathtools}
\usepackage{physics}
\usepackage{scalerel}
\usepackage{stackengine}
\usepackage{subcaption}
\usepackage{tikz}
\usepackage{xspace}
\usepackage{algorithm}
\usepackage{todonotes}
\usepackage{algpseudocode}

\usetikzlibrary{positioning, fit, calc, patterns, decorations.pathreplacing}

\renewcommand{\vec}[1]{\bm{#1}}
\newcommand{\mat}[1]{\bm{#1}}

\newcommand{\NP}{\textsc{NP}\@\xspace}
\newcommand{\QUBO}{\textsc{Qubo}\@\xspace}

\newcommand{\CIR}{\textsc{Circles}\@\xspace}
\newcommand{\CON}{\textsc{Cones}\@\xspace}

\newcommand{\BB}{\mathbb{B}}
\newcommand{\BR}{\mathbb{R}}
\newcommand{\BS}{\mathbb{S}}

\newcommand*{\defeq}{\mathrel{\vcenter{\baselineskip0.5ex\lineskiplimit0pt\hbox{\scriptsize.}\hbox{\scriptsize.}}}%
	=}

\newcommand{\HInitial}{\mat{H}_{\mathrm{I}}}
\newcommand{\HTarget}{\mat{H}_{\mathrm{P}}}

\newcommand{\SG}[1]{\gamma( #1 )}

\newcommand{\ER}[1]{\Gamma( #1 )}
\newcommand{\ratio}{\rho}

\makeatletter
\newcommand\thefontsize{The current font size is: \f@size pt}
\makeatother
\begin{document}
\title{Investigating the Relation Between Problem Hardness and QUBO Properties}
%
%
\author{Thore Gerlach\inst{1}\orcidID{0000-0001-7726-1848} \and
	Sascha M\"ucke\inst{2}\orcidID{0000-0001-8332-6169}}
\authorrunning{T. Gerlach et al.}
%
\institute{Fraunhofer IAIS, Sankt-Augustin, Germany
\email{thore.gerlach@iais.fraunhofer.de}\\
\and
Lamarr Institute, TU Dortmund University, Dortmund, Germany\\
\email{sascha.muecke@tu-dortmund.de}}
\maketitle              
\begin{abstract}
	Combinatorial optimization problems, integral to various scientific and industrial applications, often vary significantly in their complexity and computational difficulty.
	Transforming such problems into Quadratic Unconstrained Binary Optimization (\QUBO) has regained considerable research attention in recent decades due to the central role of \QUBO in Quantum Annealing.
	This work aims to shed some light on the relationship between the problems' properties.
	In particular, we examine how the spectral gap of the \QUBO formulation correlates with the original problem, since it has an impact on how efficiently it can be solved on quantum computers.
	We analyze two well-known problems from Machine Learning, namely Clustering and Support Vector Machine (SVM) training, regarding the spectral gaps of their respective \QUBO counterparts.
	An empirical evaluation provides interesting insights, showing that the spectral gap of Clustering \QUBO instances positively correlates with data separability, while for SVM \QUBO the opposite is true.
\end{abstract}
%
%
%
\section{Introduction}

Combinatorial optimization problems lie at the core of many \NP-hard problems in Machine Learning (ML)  \cite{guruswami2009hardness}:
Clustering a data set comes down to deciding, for every point, if it belongs to one cluster or another.
Training a Support Vector Machine (SVM) involves identifying the subset of support vectors (SVs).
These decisions are highly interdependent, making the tasks computationally complex.

\begin{figure}[t]
	\centering
	\resizebox{\textwidth}{!}{%
		\begin{tikzpicture}[
			main node/.style={
				draw, rounded corners=3mm,
				minimum width=3cm,
				inner sep=2mm},
			node distance=3.5cm]
			\node[main node] (n1) at (0,0) {Problem};
			\node[align=left] (n1a) [below=1mm of n1] {\scriptsize{Domain $\mathcal{S}$}\\\scriptsize{+ Data}\\\scriptsize{+ Hyperparameters}};
			\node[main node,fill=blue!5!white] (n2) [right of=n1] {\QUBO $\mat Q$};
			\node[align=left] (n2a) [below=1mm of n2] {\scriptsize{Domain $\BB^n$}};
			\node[main node] (n3) [right of=n2] {Energy $f_{\mat Q}$};
			\node[main node] (n4) [right of=n3] {Minimizer $\vec{z}^*\in\BB$};
			\node[main node] (n5) [below=1cm of n4] {Minimizer $\vec{s}^*\in\mathcal{S}$};
			\node[main node] (n6) [below=1cm of n3] {Loss function};
			\path[->,>=stealth]
			(n1) edge[bend left] node [above] {embedded in} (n2)
			(n2) edge[bend left] node [above] {induces} (n3)
			(n3) edge[bend left] node [above] {induces} (n4)
			(n1) edge[bend right=20] node[right,xshift=1cm] {induces} (n6.west)
			(n4) edge node[right] {\scriptsize corresponds} (n5)
			(n3) edge node[left,align=right] {\scriptsize preserves\\\scriptsize minimum} (n6)
			(n6) edge[bend left] node[above] {induces} (n5);
		\end{tikzpicture}
	}
	\caption{Workflow of embedding a problem into \QUBO. Every solution candidate $\vec s$ in the problem domain $\mathcal{S}$ has an easily computable corresponding $\vec z\in\BB^n$.}
	\label{fig:qubo-workflow}
\end{figure}
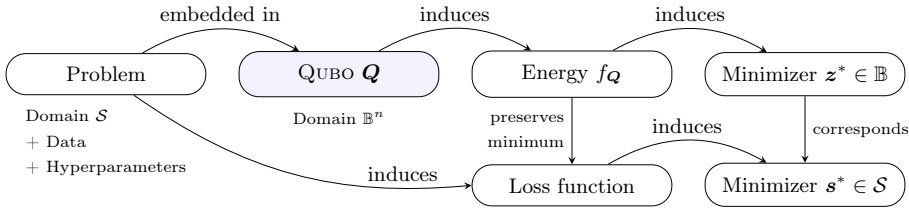

In the advent of Quantum Computing (QC), combinatorial problems have gained renewed attention due to the possibility of solving them through the exploitation of quantum tunneling effects.
Particularly, the Ising model and the equivalent \QUBO problem have become the central target problem class for Quantum Annealing (QA) \cite{Kadowaki.Nishimori.1998,Farhi.etal2000}.
In \QUBO, a parameter matrix $\mat Q$ is given that parametrizes a loss function over binary vectors, which we want to minimize.
It can be shown that, in general, this problem is \NP-hard \cite{Pardalos.Jha.1992}.
The value of \QUBO lies in its versatility:
Many \NP-hard problems can be reduced to it by means of computing $\mat Q$ from input data and hyperparameters, thus \QUBO has seen many applications in various  domains \cite{Hammer.Shlifer1971,Kochenberger.etal2005,Kochenberger.etal2014,Bauckhage.etal2018,Mucke.etal2019,piatkowski2022towards,Mucke.etal2023}.
Solving the \QUBO formulation yields a minimizing binary vector, which maps back to an optimal solution of the original problem.
\cref{fig:qubo-workflow} shows a schematic view of this workflow.

However, despite the promise of quantum speedup, not every instance of \QUBO is equally easy to solve.
It was shown that certain instances require exponential annealing time \cite{Altshuler.etal.2009}, which may render solving them on quantum computers equally infeasible as by brute force.
A central determining factor is the minimal \emph{spectral gap} (SG) of the corresponding Annealing Hamiltonian (AH), which in turn dictates the annealing speed (see \cref{sec:background}): A small SG leads to a higher probability of obtaining sub-optimal results (see e.g. \cite{Somma.etal2008}).
The SG is a physical property of the AH, and its connection to classical complexity theory is poorly understood.
One would expect that classically hard problems tend to be more difficult to solve, even using non-classical methods like QA.

We investigate this connection, both by means of an empirical study and theoretical considerations.
To this end, we take instances of optimization problems from ML, embed them into \QUBO, and compare their properties to uncover correspondences.
Our central research question is: \textbf{What is the relation between the hardness of a particular optimization problem and its corresponding \QUBO formulation?}
In the scope of this paper, ``hardness'' refers to data properties rather than complexity.
We find that the relationship surprisingly not always aligns with intuition:
With clustering, a stronger separation of data corresponds with a larger SG, while for SVM learning the opposite is true.

This paper is structured as follows:
In \cref{sec:background}, we give an overview on the background of Adiabatic QC.
\QUBO formulations for the two classical learning problems, Binary Clustering and SVM, can be found in \cref{sec:clustering,sec:svm}.
In \cref{sec:experiments}, we conduct experiments and
a conclusion is drawn in \cref{sec:conclusion}.

\section{Background}
\label{sec:background}

In a \QUBO problem, we are given an upper triangle matrix $\mat Q\in\mathbb{R}^{n\times n}$, which parameterizes the \emph{energy function} $f_{\mat Q}$ defined as \begin{equation}
	f_{\mat Q}(\vec z)\defeq \vec z^\top\mat Q\vec z=\sum_{i=1}^n\sum_{j=i}^n Q_{ij}z_iz_j\;,
\end{equation}%
where $\vec z\in\BB^n$ is a binary vector with $\BB=\lbrace 0,1\rbrace$.
The objective is to find a vector $\vec z^*$ that minimizes $f_{\mat Q}$, i.e., $\forall \vec z\in\BB^n: ~f_{\mat Q}(\vec z^*)\leq f_{\mat Q}(\vec z)$.
The Ising model is almost identical to \QUBO, but uses the binary set $\BS=\lbrace -1,+1\rbrace$.
The model's energy function can be obtained from $f_{\mat Q}$ through a simple change of $\vec z \mapsto (\vec s+1)/2$.

QA is a promising method for approximating the minimizing solutions of Ising models, first proposed by Kadowaki and Nishimori \cite{Kadowaki.Nishimori.1998}.
Instead of bits for variables it uses \emph{qubits}, which, when measured, take either state from $\BS$ with a certain probability.
Further, systems of $n$ qubits can exhibit arbitrary probability distributions over the space $\BS^n$.

The rough quantum-mechanical equivalent of loss functions in ML are \emph{Hamiltonians}, which are complex-valued hermitian matrices $\mat H\in\mathbb C^{2^n\times2^n}$ that describe the total energy of a system.
The expected energy of an $n$-qubit state $\ket{\psi}$ is given by $\bra{\psi}\mat H\ket{\psi}$, where $\ket{\psi}$ is a $2^n$-element complex vector describing the qubits' state, and $\bra{\psi}$ its conjugate transpose.
The Ising model Hamiltonian is diagonal and contains the energy values for all possible binary states $\BS^n$, which are simultaneously its eigenvalues.
The minimizing state corresponding to the smallest eigenvalue is called \emph{ground state}.
The Adiabatic theorem states that a system with a time-dependent Hamiltonian $\mat H(t)$ tends to stay in its ground state, even if the Hamiltonian slowly changes over time \cite{Born.Fock.1928}.
At the core of QA lies the idea to prepare a quantum system in the ground state of an ``easy'' Hamiltonian $\HInitial$ and slowly change it to the actual problem Hamiltonian $\HTarget$ over time: \begin{equation}\label{eq:Ht}
	\mat{H}(s) \defeq f(s)\HInitial+g(s)\HTarget\;,
\end{equation}%
with $f,g:[0,1]\to\BR_{\geq 0}$, such that $f(0)\gg g(0)$ and $f(1)\ll g(1)$.
The speed at which the Hamiltonian can safely evolve without the system leaving its ground state depends on the minimal SG, which is the minimal difference between the two lowest eigenvalues over time.
Let $\lambda_1(s),\dots,\lambda_{2^n}(s)$ denote the eigenvalues of $\mat H(s)$ in increasing order ($\lambda_i(s)<\lambda_{i+1}(s)~\forall 1\leq i<2^n$), then $\SG{\mat H(s)}\defeq \min_{s\in [0,1]} ~\lambda_{2}(s)-\lambda_{1}(s)$.
When a Hamiltonian $\mat H$ is not time-dependent, we simply write $\SG{\mat H}$ to denote the difference between its lowest eigenvalues.
A small SG requires a slow change rate, which leads to a long, potentially exponential (c.f. \cite{Altshuler.etal.2009}) \emph{annealing time}.
It is therefore desirable to somehow increase the SG by choosing $\HInitial$, $\HTarget$, $f$ and $g$ accordingly.
As $\HInitial$, $f$ and $g$ are usually prescribed by the annealing hardware at hand, the only free variable is $\HTarget$.

When talking about the SG of a \QUBO instance $\mat Q$, it is important to make the distinction between the parameter matrix $\mat Q$ and its corresponding Hamiltonian $\mat H_{\mat Q}$:
The entries along the diagonal of the latter correspond to the values $\mat z^\top\mat Q\mat z$ for every possible $\mat z\in\BB^n$.
Therefore, the SG is simply the difference between the lowest and second-to-lowest values of $f_{\mat Q}$ (which is very hard to compute for large $n$).
The eigenvalues of $\mat Q$ hold no particular relevance.

\section{QUBO Formulations and their Spectral Gaps}

The minimal SG $\SG{\mat H(s)}$ cannot be easily predicted from either $\SG{\HInitial}$ or $\SG{\HTarget}$.
However, we can still make some statements about it using known results about eigenvalues of Hermitian matrices.

\begin{theorem}[Weyl's Inequality]
	Let $\mat M, \mat N, \mat R$ be $m\times m$ Hermitian matrices with $\mat N+\mat R=\mat M$.
	Let $\mu_i,\nu_i,\rho_i$ denote their respective eigenvalues in ascending order, i.e., $\mu_i\leq\mu_{i+1}~\forall 1\leq i<m$, and for $\nu_i,\rho_i$ analogously.
	Then the following inequality holds for all $1\leq i\leq m$: \begin{equation}
		\nu_i+\rho_1 \leq \mu_i \leq \nu_i+\rho_m\;.
	\end{equation}
\end{theorem}

Applying this inequality multiple times we find the following bound on the SG of sums of two Hamiltonians: \begin{equation}
	\underbrace{\mu_2-\mu_1}_{=\SG{\mat M}}\leq\underbrace{\nu_2-\nu_1}_{=\SG{\mat N}}+ (\underbrace{\rho_m-\rho_1}_{\defeq\ER{\mat{R}}})\;.
	\label{eq:first_bound}
\end{equation}

Recall the definition of the standard time-dependent AH given in \cref{eq:Ht}.
Assume that $f,g:[0,1]\to[0,1]$, $f(0)=g(1)=1$, $f(1)=g(0)=0$, $f$ is monotonous decreasing and $g$ is monotonous increasing.
E.g., $f$ and $g$ can be chosen as $f(s)=1-s$ and $g(s)=s$ with $s=t/T_a$, where $T_a$ is the total annealing time and $t$ the current time in the annealing process.
With the assumption $f(x)=1-g(x)$, we obtain $\min_{s\in[0,1]}af(s)+bg(s)=\min_{s\in[0,1]}af(s)+b\left(1-f(s)\right)=\min\{a,b\}$,
and from \cref{eq:first_bound} follows
\begin{equation}
\SG{\mat H(s)}\le \min\{\SG{\HTarget},\SG{\HInitial}\}\le \SG{\HTarget}  \;.
\end{equation}
%

This result provides motivation to increase $\SG{\HTarget}$ when trying to improve QA performance, as it is an upper bound on $\SG{\mat H(s)}$:
Increasing it does not guarantee a larger minimal SG, but is a necessary precondition.

\subsection{Kernel 2-Means Clustering}\label{sec:clustering}

Our first \QUBO embedding of interest is clustering:
Assume we are given a set of $n$ data points $\mathcal X\subset \BR^d$, $\left|\mathcal X\right|=n$.
We want to partition $\mathcal{X}$ into disjoint clusters $\mathcal X_1,\mathcal{X}_2\subset\mathcal X$, $\mathcal X_1\dot{\cup}\mathcal X_2=\mathcal X$.
We gather the data in a matrix $\mat{X}\defeq\left[\vec{x}^1,\dots,\vec{x}^n\right], \forall i: \vec x^i\in\mathcal{X}$, and assume that it is centered, i.e., $\mat{X}\vec{1}=\vec{0}$, where $\vec{1}$ denotes the $n$-dimensional vector consisting only of ones. 
A \QUBO formulation was derived in \cite{Bauckhage.etal2018}, which minimizes the \textit{within cluster scatter}:
\begin{equation}
	\min_{\vec{s}\in\BS^n}-\vec{s}^{\top}\mat{X}^{\top}\mat{X}\vec{s}
	\Leftrightarrow\min_{\vec{z}\in\BB^n}-\vec{z}^{\top}\mat{X}^{\top}\mat{X}\vec{z}+\vec{1}^{\top}\mat{X}^{\top}\mat{X}\vec{z}\;,
	\label{eq:clustering_ising}
\end{equation}
where $\vec{s}=2\vec{z}-\vec{1}$.
A value $z_i=1$ indicates that data point $\vec{x}^i$ is in cluster $\mathcal X_1$, and in $\mathcal X_2$ for $z_i=0$.
Observing that $\mat{X}^{\top}\mat{X}$ is a Gram matrix leads to a possible application of the kernel trick. 
For this, we consider a centered kernel matrix $\mat{K}\in\mathbb R^{n\times n}$ with elements $k(\vec{x}^i,\vec{x}^j)$,
where $k:\mathbb R^n\times\mathbb R^n\to \mathbb R$ is a kernel function.
$k(\vec{x}^i,\vec{x}^j)$ indicates how similar data points $\vec{x}^i$ and $\vec{x}^j$ are in some feature space.
We can reformulate \cref{eq:clustering_ising} to
\begin{align}
	\min_{\vec{z}\in\BB^n}\vec{1}^{\top}\mat{K}\vec{z} -\vec{z}^{\top}\mat{K}\vec{z}
	\Leftrightarrow\min_{\vec{z}\in\BB^n}\sum_{i,j=1}^nK_{ij}(1-z_i)z_j  
	\Leftrightarrow\min_{\mathcal X_1,\mathcal X_2}\sum_{\vec{x}\in\mathcal X_1,\vec{y}\in\mathcal X_2}k(\vec{x},\vec{y}).
	\label{eq:clustering_qubo}
\end{align}
We can attribute problem properties to effects on the SG of the resulting Hamiltonian of the \QUBO formulation, by observing that the similarities between the different clusters are summed up.
We see that the \QUBO energy according to \cref{eq:clustering_qubo} is negatively correlated to the similarities within the clusters and positively correlated to the similarities between the clusters.

Thus, we claim that the SG of the \QUBO formulation in \cref{eq:clustering_qubo} is \begin{enumerate*}[label=(\roman*)]\item positively correlated to the \textit{separability} (the inter-cluster distance), and \item negatively correlated to the \textit{compactness} (the intra-cluster distances),\end{enumerate*} which we validate in \cref{sec:experiments:clustering}.

\subsection{Simple Support Vector Machine Embedding}\label{sec:svm}

A linear Support Vector Machine (SVM) is a classifier that takes a labeled data set $\mathcal{D}=\lbrace (\vec x^i, y_i)\rbrace$ with $\vec x^i\in\mathbb{R}^d$ and $y_i\in\lbrace -1,+1\rbrace$ for $i\in[n]\defeq\{1,\dots,n\}$, and separates them with a hyperplane \cite{Cortes.Vapnik.1995}.
As there may be infinitely many such hyperplanes, an additional objective is to maximize the \emph{margin}, which is the area around the hyperplane containing no data points, in order to obtain best generalization.
The hyperplane is represented as a normal vector $\bm{w}\in\mathbb{R}^d$ and an offset or \emph{bias} $b\in\mathbb{R}$.
To ensure correctness, the optimization is subject to  $\langle\bm{w},\bm{x}^i\rangle-b)\cdot y_i\geq 1-\xi_i$, i.e., every data point must lies on the correct side of the plane.
As for real-world data perfect linear separability is unlikely, \emph{slack variables} $\xi_i>0$ allow for slight violations, which we want to minimize.
This yields a primal objective function of \begin{align*}
	\text{minimize} &~\frac{1}{2}\lVert\bm{w}\rVert^2_2+C\sum_i\xi_i\\
	\text{s.t.} &~\forall i. ~(\langle\bm{w},\bm{x}^i\rangle-b)\cdot y_i\geq 1-\xi_i\;.
\end{align*}
We optimize over $\bm{w}$, $b$ and $\vec \xi$, while $C>0$ is a hyperparameter controlling the impact of misclassification.
Typically this problem is solved using its well-established Lagrangian dual \begin{align}
	\text{maximize} &~\sum_{i=1}^n \alpha_i -\frac{1}{2}\sum_{i=1}^n\sum_{j=1}^n\alpha_i\alpha_jy_iy_j\langle\bm{x}^i,\bm{x}^j\rangle \label{eq:svmdual}\\
	\text{s.t.} &~\sum_i\alpha_iy_i=0, \quad 0\leq\alpha_i\leq C ~\forall i\;. \label{eq:kkt}
\end{align}

Following \cite{Mucke.etal2019} we make the simplifying assumption that $\alpha_i$ can only take the values $0$ or $C$, which allows us to introduce binary variables $z_i\in\BB$ and write $\alpha_i = Cz_i ~\forall i$.
The condition \cref{eq:kkt} can be included in the main objective by introducing the penalty term $-\lambda\left(\sum_i\alpha_iy_i\right)^2$, which is 0 when the condition is fulfilled, and negative otherwise.
Similarly to the Clustering case (see \cref{sec:clustering}), we can apply the kernel trick and derive the following \QUBO formulation:
\begin{align}
&\min_{\vec{z}\in\{0,1\}^n}-\vec{1}^{\top}\vec{z}+C\vec{z}^{\top}\left(\frac{1}{2}\left(\mat{Y}\odot\mat{K}\right)+\lambda \mat{Y}\right)\vec{z} 
\label{eq:svm_qubo}\;,
\end{align}%
where 
$\mat Y$ has entries $Y_{ij}=y_iy_j ~\forall i,j\in[n]$ and $\odot$ denotes the entry-wise product.
The parameter $\lambda$ has to be chosen large enough to ensure \cref{eq:kkt} is fulfilled, however, if it is much larger than the objective in \cref{eq:svmdual}, the SG will be very small \cite{benkner2020adiabatic}.
$C$ controls how ``soft'' the margin is, i.e., how strongly misclassified data points are penalized.
A large $C$ does so heavily, which may result in overfitting.

We claim that the SG is negatively correlated \begin{enumerate*}[label=(\roman*)]\item with $C$, \item with $\lambda$, and also \item with the \textit{separability}, i.e., actual margin size of the data \end{enumerate*}.
We validate this in \cref{sec:experiments:svm}.

\section{Experiments}\label{sec:experiments}

At the core of this work we conduct an empirical evaluation of \QUBO formulations and their properties.
For each experiment, the steps we take are as follows:
\begin{enumerate*}
	\item Choose \textbf{problem type} and \textbf{hyperparameters},
	\item Sample \textbf{data set} with known properties,
	\item Compute \textbf{QUBO parameters} and record the SG.
\end{enumerate*}
Using the acquired data, we investigate the relationship between data parameters and SG, which has a high impact on problem hardness for QC, as we have shown before.
We consider the two problems of binary clustering and SVM training, which we described in previous sections.
As input data, we sample synthetic data sets for each repetition of the experiments. 
To this end, we consider two different types, \CON and \CIR, both of which have parameters allowing us to vary the resulting optimization problems' difficulty by adjusting the data class separation.

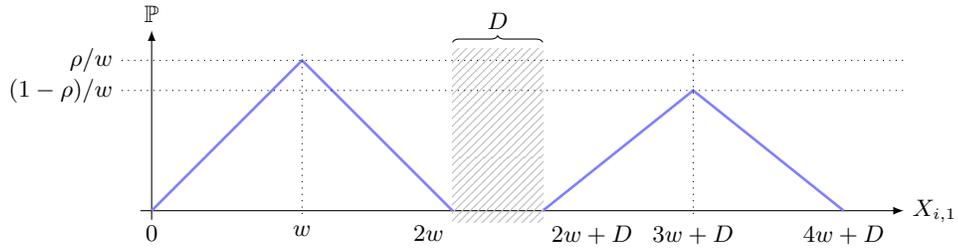
\begin{figure}[t!]
	\centering
	\begin{tikzpicture}[xscale=2, yscale=0.8,graph/.style={line width=1.0, draw=blue!50!white}]
		\draw[->,>=latex] (-0.075,0) -- (5,0) node[anchor=west] {$X_{i,1}$};
		\draw[->,>=latex] (0,-0.15) -- (0,3) node[anchor=south] {$\mathbb{P}$};
		\draw[graph] (0, 0) -- (1, 2.5) -- (2, 0);
		\draw[graph] (2.6, 0) -- (3.6, 2) -- (4.6, 0);
		\draw[pattern=north east lines,pattern color=gray!50!white,draw=none] (2,-0.2) rectangle (2.6,2.7);
		\draw[decorate,decoration=brace] (2,2.8) --  (2.6,2.8) node[midway,anchor=south,yshift=0.6mm] {$D$};
		\draw[dotted] (-0.2,2.5) node[anchor=east] {$\ratio/w$} -- (5,2.5);
		\draw[dotted] (-0.2,2) node[anchor=east] {$(1-\ratio)/w$} -- (5,2);
		\draw[dotted] (1, 2.6) -- (1, -0.1) node[anchor=north] {$w$};
		\draw[dotted] (3.6, 2.6) -- (3.6, -0.1) node[anchor=north] {$3w+D$};
		\node[anchor=north] at (0, -0.1) {$0$};
		\node[anchor=north east] at (2, -0.1) {$2w$};
		\node[anchor=north west] at (2.6, -0.1) {$2w+D$};
		\node[anchor=north] at (4.6, -0.1) {$4w+D$};
	\end{tikzpicture}
	\caption{Distribution of the first dimension of the 2-dimensional synthetic data used for our experiments (before applying the rotation): Two clusters are sampled such that there is a separating margin of at least size $D$ between them. The parameter $w$ controls the spread of data points, while $r$ is the ratio between the number of data points in the first vs. the second cluster.}
	\label{fig:datadistr}
\end{figure}

\begin{figure}[t!]
	\centering
	\includegraphics[width=\textwidth]{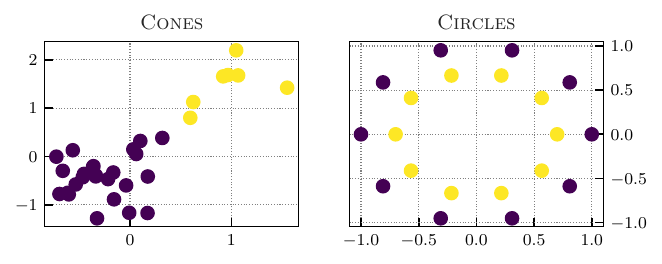}
	\caption{Exemplary instances of the data sets used for our experiments.}
	\label{fig:data}
\end{figure}

\subsubsection{\CON}

Let $n\in\mathbb{N}$ with $n\geq 2$ denote the number of data points, $\ratio\in(0,1)$ the cluster size ratio, $w>0$ a spread parameter, and $D\in\BR$ a separating margin size.
We set $n_1\defeq\min\lbrace 1,\lfloor \ratio n\rfloor\rbrace$ and $n_2\defeq 1-n_1$ as the cluster sizes.
We create a matrix $\mat{X}\in\BR^{n\times 2}$ where every entry $X_{ij}$ is sampled i.i.d. from a triangular distribution within the interval $[0, 2w]$ and with mode $w$.
We chose the triangular distribution over a normal distribution because it has no outliers, which allows us to define a hard lower bound on the separating margin between clusters.
For all $i>n_1$ we then set $X_{i,1}\mapsto X_{i,1}+2w+D$, which shifts all of these points such that $\norm{\vec X_{i,\cdot}-\vec X_{\ell,\cdot}}_2\geq D$ is tight for all $i\in[n_1]$ and $n_1<\ell\leq n$.
The distribution of $\vec X_{\cdot,1}$ is visualized in \cref{fig:datadistr}; the distribution of $\vec X_{\cdot,2}$ consists of just a single triangle from $0$ to $2w$ with height $1/w$ at mode $w$.
Next, we sample $\theta$ uniformly from $[0, 2\pi)$ and apply $\mat X\mapsto\mat X\mat R(\theta)$, where $\mat R(\theta)$ is a 2D rotation matrix.
This rotation leaves the distances unchanged but introduces another degree of freedom.
Lastly, we center the data by computing $\vec\mu_j\defeq\sum_{i=1}^nX_{ij}/n$ and applying $X_{ij}\mapsto X_{ij}-\vec\mu_j$ for all $i,j\in [n]\times [2]$.
The target vector $\vec y\in\mathbb{S}^n$ is set to $y_i=-1$ for $i\in [n_1]$ and $y_i=+1$ for $n_1<i\leq n$.

\subsubsection{\CIR}

As a second data set type, we consider two circles, which are not linearly separable in $\mathbb R^2$.
The radius of the outer circle is fixed to 1 and for our experiments we vary the radius of the inner circle $r$.
The circles consist of an equal number of points $n/2$ and Gaussian noise with standard deviation $\sigma$ is added to every point (see \cref{fig:data}, right).
To bridge the gap to linear separability, we project the data set to a higher-dimensional feature space via a feature map $\phi:\mathbb R^2\to \mathbb R^3$, $\phi(\vec{x})\defeq (x_1, x_2, a \left\|\vec{x}\right\|^2)^{\top}$.
In this space, the data is linearly separable.
A corresponding kernel function is given by $k(\vec{x},\vec{y})\defeq \langle\phi(\vec{x}),\phi(\vec{y})\rangle= \langle\vec{x},\vec{y}\rangle+a^2\norm{\vec{x}}^2\norm{\vec{y}}^2$.
For every experiment, we compare three different problem sizes, that is, we consider $n\in\{8,20,32\}$.
To make SG comparable between \QUBO instances of the same size, we scale each $\mat Q$ such that $\norm{\mat Q}_{\infty}=1$.

\subsection{Clustering}\label{sec:experiments:clustering}

\begin{figure}
	\centering
	\includegraphics[width=\textwidth]{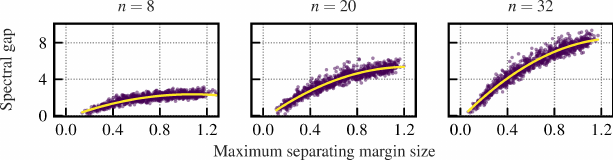}
	\caption{Spectral gap of \QUBO instances according to \cref{eq:clustering_qubo} against maximum separating margin size $D$ for \CON; $w=0.2,~\ratio=0.5$ fixed, 1000 random data sets with $n\in\{8,20,32\}$ and $D\in[0, 1]$ uniformly sampled. The yellow curve is a fitted quadratic function.}
	\label{fig:clustering_cones_margin}
\end{figure}

\begin{figure}
	\centering
	\includegraphics[width=\textwidth]{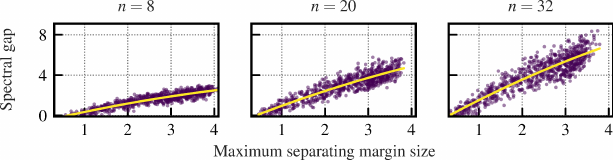}
	\caption{Spectral gap of \QUBO instances according to \cref{eq:clustering_qubo} against maximum separating margin size $D$ for \CIR; $\sigma=0.05, ~\ratio=0.5$ fixed, 1000 random data sets with $n\in\{8,20,32\}$ and $r\in[0, 1]$ uniformly sampled. The yellow curve is a fitted quadratic function.}
	\label{fig:clustering_circles_margin}
\end{figure}

We first explore the Clustering \QUBO in \cref{eq:clustering_qubo}. 
Changing the maximum separating margin size between the two clusters changes the SG of the corresponding \QUBO instance, as shown in \cref{fig:clustering_cones_margin,fig:clustering_circles_margin}:
For \cref{fig:clustering_cones_margin}, we sample $1000$ different \CON data sets with varying cluster distances in $[0,1]$ and fix $w=0.2$, $\ratio=0.5$. 
We find that the SG is increasing with an increasing problem size $n$ and that there is a clear quadratic positive correlation between the SG and the margin size.

A similar setup can be found in \cref{fig:clustering_circles_margin}, where $1000$ different \CIR data sets are sampled with varying inner radius in $[0,1]$ and $\sigma=0.05$, $\ratio=0.5$. 
Again we find that the SG increases with an increasing margin size, but with a linear correlation.
Since different kernels are used in \cref{fig:clustering_cones_margin} and \cref{fig:clustering_circles_margin}, the exact correlation form is dependent on the exact data set and the used kernel.

\begin{figure}
	\centering
	\includegraphics[width=\textwidth]{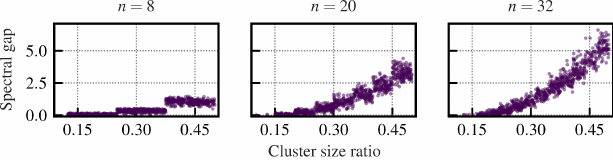}
	\caption{Spectral gap of \QUBO instances according to \cref{eq:clustering_qubo} against cluster ratio for \CON; $D=0.5, ~w=0.2$ fixed, 1000 random data sets for $n\in\{8,20,32\}$ and $\ratio\in[0.1, 0.5]$ uniformly sampled. }
	\label{fig:clustering_cones_ratio}
\end{figure}

In \cref{fig:clustering_cones_ratio}, the effects of a varying cluster ratio in $\ratio\in[0.1,0.5]$ are depicted for the \CON setup.
We again sample 1000 data sets with fixing $w=0.2$ and $D=0.5$.
A positive correlation becomes evident between cluster ratio and SG. 
That is, the \QUBO problem is easier to solve with QC when the clusters have the same size.
The effect that the plots look like a step function for small $n$ is due to the fact that there $n/2$ different configurations, e.g., for $n=8$, we can have the four cases $n_1=1$, $n_1=2$, $n_1=3$ and $n_1=4$.

\begin{figure}[t!]
	\centering
	\includegraphics[width=\textwidth]{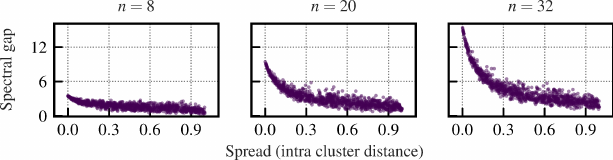}
	\caption{Spectral gap of \QUBO instances according to \cref{eq:clustering_qubo} against spread for \CON; $D=0.5,~\rho=0.5$ fixed, 1000 random data sets for $n\in\{8,20,32\}$ and $w\in[0, 1]$ uniformly sampled.}
	\label{fig:clustering_cones_noise}
\end{figure}

In \cref{fig:clustering_cones_noise}, we vary the spread $w\in[0,1]$ for 1000 different data sets and fix $D=0.5$, $\rho=0.5$.
We can see that the SG is negatively correlated to the spread of the data set.

Putting the results together we can deduce that the is SG positively correlated with the inter-cluster distance (separability) and negatively correlated with the intra-cluster distance, supporting our claims in \cref{sec:clustering}.

\subsection{Support Vector Machine}\label{sec:experiments:svm}

\begin{figure}[t!]
	\centering
	\includegraphics[width=\textwidth]{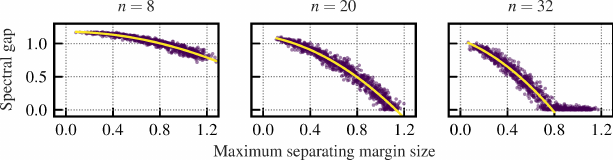}
	\caption{Spectral gap of \QUBO instances according to \cref{eq:svm_qubo} against maximum separating margin size $D$ for \CON. Same configuration as for \cref{fig:clustering_cones_margin}.
}
	\label{fig:svm_cones_margin}
\end{figure}

\begin{figure}[t]
	\centering
	\includegraphics[width=\textwidth]{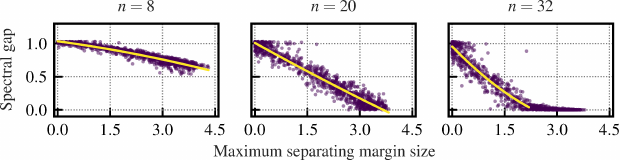}
	\caption{Spectral gap of \QUBO instances according to \cref{eq:svm_qubo} against maximum separating margin size $D$ for \CIR. Same configuration as for \cref{fig:clustering_circles_margin}.}
	\label{fig:svm_circles_margin}
\end{figure}

We move over to experiments with the SVM \QUBO in \cref{sec:svm}.
Again, we depict the effect of changing the maximum separating margin size between the two clusters on the SG of the corresponding \QUBO in \cref{fig:svm_cones_margin,fig:svm_circles_margin}.
We use the same parameters for the data sets as in \cref{fig:clustering_cones_margin,fig:clustering_circles_margin}.
Interestingly, we now observe a negative correlation between the SG and the margin size, making the problem harder to solve with a quantum computer if the data is well separable.
For $n=32$ the spectral basically vanishes from a certain margin size for \CIR.
Furthermore, we again observe that this correlation is quadratic for \CON and linear for \CIR, leading to a large dependence on the used kernel function and the data set at hand.

Note that we are considering a \QUBO formulation for a soft-margin SVM: even though the SG might be very small, the second best solution might also be satisfactory for solving the original problem.
In contrast, the second best solution of a Clustering \QUBO is much worse: changing a single bit leads to a data point within the wrong cluster, which increases the energy more dramatically the further the two clusters are separated.

\begin{figure}[t]
	\centering
	\includegraphics[width=\textwidth]{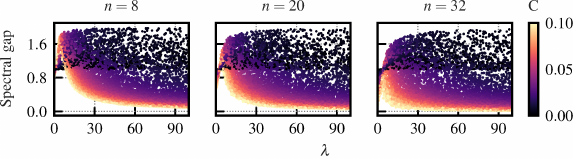}
	\caption{Spectral gap of \QUBO instances according to \cref{eq:svm_qubo} against $\lambda$ and $C$ for \CON; $w=0.2,~\ratio=0.5$ fixed, $10\,000$ random data sets for $n\in\{8,20,32\}$ and $\lambda\in[0, 100],~C\in[0,0.1]$ uniformly sampled.
	}
	\label{fig:svm_cones_C_kkt_100}
\end{figure}

\begin{figure}[t]
	\centering
	\includegraphics[width=\textwidth]{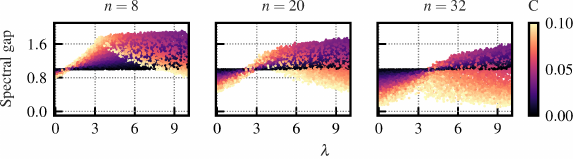}
	\caption{Same as \cref{fig:svm_cones_C_kkt_100}, zoomed in on $\lambda\in [0,10]$.}
	\label{fig:svm_cones_C_kkt_10}
\end{figure}

In \cref{fig:svm_cones_C_kkt_100,fig:svm_cones_C_kkt_10}, we show the effect of varying $\lambda$ and $C$ on the SG for \CON.
We fix $\ratio=0.5$, $w=0.2$, $D=0.5$ and sample $10\,000$ data sets with $C\in[0,0.1]$, $\lambda\in[0,100]$ for \cref{fig:svm_cones_C_kkt_100} and $\lambda\in[0,10]$ for \cref{fig:svm_cones_C_kkt_10}, respectively.
It is evident from \cref{fig:svm_cones_C_kkt_100} that the SG decreases as $\lambda$ and $C$ are increased.
However, there are interesting intervals for $\lambda$ when fixing $C$, such that the SG first increases and then decreases with increasing $\lambda$, forming a triangular shape when plotted, which gets more pointy with an increasing value of $C$ -- see \cref{fig:svm_cones_C_kkt_10} for a closer view.
We observe a similar effect with \CIR.

Combining our observations we deduce that the SG is negatively correlated with the inter-cluster distance (separability) and the parameters $\lambda$, $C$ except for a small region, supporting our claims from \cref{sec:svm}.

\section{Conclusion}\label{sec:conclusion}

In this paper, we investigated the connection between the problem hardness of classical ML problems and their solvability on quantum hardware.
We considered \QUBO formulations for the $2$-Means Clustering and SVM learning.
We highlighted that the SG of these formulations impact their solvability on quantum hardware, and showed how the SG behaves when adapting the problem parameters, which we underpinned with an empirical study.

We found that for $2$-Means Clustering an easier problem also leads to a better solvability on quantum computers. 
Here, ``easy'' refers to the separability and compactness of the different classes. 
We found a positive correlation between these properties and the SG of the corresponding \QUBO. 
Interestingly, this is not the case for the SVM problem, where we would expect a better solvability of a problem with a large separation of the classes. 
However, we found a negative correlation between separation and SG.
Furthermore, other hyperparameters, such as the one controlling the softness of the margin avoiding overfitting, are negatively correlated to the SG.
This is due to the balancing different objectives in one single \QUBO problem.
Combining these two insights, we conclude that the original problems' hardness is not directly connected to the solvability on quantum computers, as one might assume.
Instead, it depends not only on the data set at hand, but also on specifics in the used \QUBO formulation. 
It would be insightful to compare the properties of more \QUBO formulations of more problems in future work.
Furthermore, investigating the effect of the \QUBO parameters and not only the problem parameters is an interesting research direction \cite{mucke2023optimum}.
This could strengthen our intuition about which problems are hard for quantum computers in particular, and the potential and limitations of QC in general.

\bibliographystyle{splncs04}

\end{document}